\documentstyle[12pt,psfig]{article}
\makeatletter
\def\@cite#1#2{{[{#1}]\if@tempswa\typeout
{IJCGA warning: optional citation argument
ignored: `#2'} \fi}}

\newcount\@tempcntc
\def\@citex[#1]#2{\if@filesw\immediate\write\@auxout{\string\citation{#2}}\fi
  \@tempcnta\z@\@tempcntb\m@ne\def\@citea{}\@cite{\@for\@citeb:=#2\do
    {\@ifundefined
       {b@\@citeb}{\@citeo\@tempcntb\m@ne\@citea\def\@citea{,}{\bf ?}\@warning
       {Citation `\@citeb' on page \thepage \space undefined}}%
    {\setbox\z@\hbox{\global\@tempcntc0\csname b@\@citeb\endcsname\relax}%
     \ifnum\@tempcntc=\z@ \@citeo\@tempcntb\m@ne
       \@citea\def\@citea{,}\hbox{\csname b@\@citeb\endcsname}%
     \else
      \advance\@tempcntb\@ne
      \ifnum\@tempcntb=\@tempcntc
      \else\advance\@tempcntb\m@ne\@citeo
      \@tempcnta\@tempcntc\@tempcntb\@tempcntc\fi\fi}}\@citeo}{#1}}
\def\@citeo{\ifnum\@tempcnta>\@tempcntb\else\@citea\def\@citea{,}%
  \ifnum\@tempcnta=\@tempcntb\the\@tempcnta\else
   {\advance\@tempcnta\@ne\ifnum\@tempcnta=\@tempcntb \else \def\@citea{--}\fi
    \advance\@tempcnta\m@ne\the\@tempcnta\@citea\the\@tempcntb}\fi\fi}
\makeatother
\newenvironment{Eqnarray}%
     {\arraycolsep 0.14em\begin{eqnarray}}{\end{eqnarray}}

\def\simlt{\stackrel{<}{{}_\sim}}
\def\simgt{\stackrel{>}{{}_\sim}}
\def\be{\begin{equation}}
\def\ee{\end{equation}}
\def\bear{\be\begin{array}}
\def\eear{\end{array}\ee}
\def\bea{\begin{Eqnarray}}
\def\eea{\end{Eqnarray}}

\def\lsim{\mathrel{\raise.3ex\hbox{$<$\kern-.75em\lower1ex\hbox{$\sim$}}}}
\def\gsim{\mathrel{\raise.3ex\hbox{$>$\kern-.75em\lower1ex\hbox{$\sim$}}}}
\def\ifmath#1{\relax\ifmmode #1\else $#1$\fi}
\def\ls#1{\ifmath{_{\lower1.5pt\hbox{$\scriptstyle #1$}}}}

\def\beq{\begin{equation}}
\def\eeq{\end{equation}}
\def\beqa{\begin{Eqnarray}}
\def\eeqa{\end{Eqnarray}}

\def\baselinestretch{1}
\begin{document}
\def\IJMPA #1 #2 #3 {{\sl Int.~J.~Mod.~Phys.}~{\bf A#1}\ (19#2) #3$\,$}
\def\MPLA #1 #2 #3 {{\sl Mod.~Phys.~Lett.}~{\bf A#1}\ (19#2) #3$\,$}
\def\NPB #1 #2 #3 {{\sl Nucl.~Phys.}~{\bf B#1}\ (19#2) #3$\,$}
\def\PLB #1 #2 #3 {{\sl Phys.~Lett.}~{\bf B#1}\ (19#2) #3$\,$}
\def\PR #1 #2 #3 {{\sl Phys.~Rep.}~{\bf#1}\ (19#2) #3$\,$}
\def\JHEP #1 #2 #3 {{\sl JHEP}~{\bf #1}~(19#2)~#3$\,$}
\def\PRD #1 #2 #3 {{\sl Phys.~Rev.}~{\bf D#1}\ (19#2) #3$\,$}
\def\PTP #1 #2 #3 {{\sl Prog.~Theor.~Phys.}~{\bf #1}\ (19#2) #3$\,$}
\def\PRL #1 #2 #3 {{\sl Phys.~Rev.~Lett.}~{\bf#1}\ (19#2) #3$\,$}
\def\RMP #1 #2 #3 {{\sl Rev.~Mod.~Phys.}~{\bf#1}\ (19#2) #3$\,$}
\def\ZPC #1 #2 #3 {{\sl Z.~Phys.}~{\bf C#1}\ (19#2) #3$\,$}
\def\PPNP#1 #2 #3 {{\sl Prog. Part. Nucl. Phys. }{\bf #1} (#2) #3$\,$}

\catcode`@=11
\newtoks\@stequation
\def\subequations{\refstepcounter{equation}%
\edef\@savedequation{\the\c@equation}%
  \@stequation=\expandafter{\theequation}
  \edef\@savedtheequation{\the\@stequation}
  \edef\oldtheequation{\theequation}%
  \setcounter{equation}{0}%
  \def\theequation{\oldtheequation\alph{equation}}}
\def\endsubequations{\setcounter{equation}{\@savedequation}%
  \@stequation=\expandafter{\@savedtheequation}%
  \edef\theequation{\the\@stequation}\global\@ignoretrue

\noindent}
\catcode`@=12
\begin{titlepage}

\title{{\bf Theoretical Constraints on the \\
Vacuum Oscillation Solution to the \\
Solar Neutrino Problem }}
\vskip2in
\author{ 
{\bf J.A. Casas$^{1,2}$\footnote{\baselineskip=16pt E-mail: {\tt
casas@mail.cern.ch}}}, 
{\bf J.R. Espinosa$^{2}$\footnote{\baselineskip=16pt E-mail: {\tt
espinosa@mail.cern.ch}. }
\footnote{\baselineskip=16pt
On leave of absence from Instituto de Matem\'aticas
y F\'{\i}sica Fundamental, CSIC, Madrid (Spain)}}, 
{\bf A. Ibarra$^{1}$\footnote{\baselineskip=16pt  E-mail: {\tt
alejandro@makoki.iem.csic.es}}} and 
{\bf I. Navarro$^{1}$\footnote{\baselineskip=16pt E-mail: {\tt
ignacio@makoki.iem.csic.es}}}\\ 
\hspace{3cm}\\
 $^{1}$~{\small Instituto de Estructura de la materia, CSIC}\\
 {\small Serrano 123, 28006 Madrid}
\hspace{0.3cm}\\
 $^{2}$~{\small Theory Division, CERN}\\
{\small CH-1211 Geneva 23, Switzerland}.
}
\date{}
\maketitle
\def\baselinestretch{1.15}
\begin{abstract}
\noindent
The vacuum oscillation (VO) solution to the solar anomaly requires an
extremely small neutrino mass splitting, $\Delta m^2_{sol}\simlt
10^{-10}\ {\rm eV}^2$.  We study under which circumstances this
small splitting (whatever its origin)  is or is not spoiled by radiative
corrections. The
results depend dramatically on the type of neutrino spectrum. If
$m_1^2\sim m_2^2\simgt m_3^2$, radiative corrections always induce
too large mass splittings. 
Moreover, if $m_1$ and $m_2$ have equal signs, the
solar mixing angle is driven by the renormalization group evolution
to very small values,
incompatible with the VO scenario (however, the results could be 
consistent with the small-angle MSW scenario).  If $m_1$ and $m_2$ have
opposite signs, the results are analogous, except for some small
(though interesting) windows in which the VO solution may be natural with
moderate fine-tuning. Finally, for a hierarchical
spectrum of neutrinos,  $m_1^2\ll m_2^2\ll m_3^2$, radiative
corrections are not dangerous, and therefore this scenario is
the only plausible one for the VO solution.
 \end{abstract}

\thispagestyle{empty}
\leftline{}
\leftline{CERN-TH/99-171}
\leftline{June 1999}
\leftline{}

\vskip-23cm
\rightline{}
\rightline{IEM-FT-195/99}
\rightline{CERN-TH/99-171}
\rightline{IFT-UAM/CSIC-99-23}
\rightline{hep-ph/9906281}
\vskip3in

\end{titlepage}
\setcounter{footnote}{1} \setcounter{page}{1}
\newpage
\baselineskip=20pt

\noindent

\section{Introduction}

There are three main explanations of the
solar neutrino flux deficits, requiring oscillations of electron neutrinos
into other species. Namely, the small and large angle MSW solutions, and
the vacuum oscillation (VO) solution. In this paper we focus on the
latter, which requires the relevant mass splitting and  mixing angle 
in the range \cite{range}
\bea 
5\times10^{-11}\ {\rm eV}^2 <  &\Delta m^2_{sol}& <
1.1\times10^{-10}\ {\rm eV}^2, \nonumber\\ 
\sin^22\theta_{sol} &>& 0.67\, .
\label{VO}
\eea
On the other hand, Superkamiokande observations \cite{SK} of
atmospheric neutrinos require neutrino oscillations 
(more precisely $\nu_\mu-\nu_\tau$ oscillations if we do not consider
oscillations into sterile species)
driven by a mass splitting and a mixing angle in the range
\cite{range}
\bea 5\times10^{-4}\ {\rm eV}^2 <  &\Delta m^2_{atm}& < 10^{-2}\ {\rm
eV}^2\ , \nonumber\\ 
\sin^22\theta_{atm}&>&0.82\ .
\label{atm}
\eea
Let us remark the enormous hierarchy of  mass splittings\footnote{It has
been pointed out~\cite{CL} that disregarding Cl data on solar
neutrinos, vacuum oscillations with much larger mass splitting could account 
for the solar anomaly.} between the different
species of neutrinos, $\Delta m^2_{sol}\ll\Delta m^2_{atm}$, which is
apparent from eqs.(\ref{VO}, \ref{atm}). 

It has been argued that the extreme tinyness of $\Delta m^2_{sol}$ in
this scenario could be related to some continuous or discrete
symmetry at high energy \cite{sym}. However, independently of the
origin of the small splittings, it must be required that their size is
not spoiled by radiative corrections, the dominant part of which can be
accounted by integrating the renormalization group equations (RGEs) between
the scale at which
the effective mass matrix is generated  and low energy. The aim of
this paper is to analyze under which  circumstances this is in fact
the case. As a result, we obtain important theoretical
restrictions on the VO scenario.

\vspace{0.2cm}
\noindent
Let us introduce now some notation. We define the effective mass term for the
three light (left-handed) neutrinos in the flavour basis,
$\nu^T=(\nu_e,\nu_\mu,\nu_\tau)$, as
\bea {\cal L}=-\frac{1}{2} \nu^T {\cal M_\nu}  \nu\;+\;{\rm h.c.}
\label{Mnu}
\eea
The mass matrix, ${\cal M_\nu}$, is diagonalized in the usual way, i.e.
${\cal M_\nu} = V^* D\, V^\dagger$, where $D=
{\rm diag}(m_1e^{i\phi},m_2e^{i\phi'},m_3)$ and
$V$ is a unitary `CKM' matrix,
relating  flavour to mass eigenstates
\bea 
\pmatrix{\nu_e \cr \nu_\mu\cr
\nu_\tau\cr}= \pmatrix{c_2c_3 &   c_2s_3 &   s_2e^{-i\delta}\cr
-c_1s_3-s_1s_2c_3e^{i\delta} &   c_1c_3-s_1s_2s_3e^{i\delta} &
s_1c_2\cr s_1s_3-c_1s_2c_3e^{i\delta} &   -s_1c_3-c_1s_2s_3e^{i\delta}
&   c_1c_2\cr}\, \pmatrix{\nu_1\cr \nu_2\cr \nu_3\cr}\,.
\label{CKM}
\eea 
Here, $s_i$ and $c_i$ denote $\sin\theta_i$ and $\cos\theta_i$,
respectively, and in the rest of the paper we will neglect CP-violating
phases. In the following, we label the mass eigenstates
$\nu_i$ in such a way that 
$|\Delta m^2_{12}|<|\Delta m^2_{23}|\sim |\Delta m^2_{13}|$, 
where $\Delta m^2_{ij}\equiv
m^2_{j}-m^2_{i}$ ($m_{\nu_3}^2$ is thus the most split eigenvalue).
Consequently, $\Delta m^2_{sol}$, $\theta_{sol}$
correspond to $\Delta m^2_{12}$, $\theta_{3}$, while
$\Delta m^2_{atm}$, $\theta_{atm}$
correspond to $\Delta m^2_{23}\sim \Delta m^2_{13}$, $\theta_{1}$
respectively. In our notation $\Delta m^2_{sol}$, $\Delta m^2_{atm}$
denote always the ``experimental'' splittings of eqs.(\ref{VO}, \ref{atm}),
while $\Delta m^2_{12}$, $\Delta m^2_{23}$ denote the computed splittings
once the radiative corrections are incorporated.
Concerning the $\theta_2$ angle,  according to the most recent 
combined analysis of SK $+$ CHOOZ data (last paper of ref.~\cite{range})
it is constrained to have low values,
$\sin^22\theta_2<0.36\, (0.64)$ at 90\% (99\%) C.L.

We assume along the paper that the effective
mass matrix for the left-handed neutrinos, ${\cal M}_\nu$, is
generated at some high energy scale, $\Lambda$, by some unspecified
mechanism. Below $\Lambda$, we consider two possibilities: either
the effective theory is the Standard Model (SM) or the minimal  
supersymmetric Standard Model (MSSM) with unbroken $R-$parity. In the
first case, the lowest dimension operator producing a
mass term of this kind is uniquely given by \cite{eff}
\bea -\frac{1}{4}\kappa \nu^T   \nu H H \;+\;{\rm h.c.}
\label{kappa}
\eea
where $\kappa$ is a matricial coupling and $H$ is the ordinary
(neutral) Higgs. Obviously, ${\cal
M}_\nu=\frac{1}{2}\kappa \langle H\rangle^2$. The effective coupling
$\kappa$ runs with the scale from $\Lambda$ to $M_Z$, with a RGE given by
\cite{Babu}
\bea 16\pi^2 \frac{d \kappa}{dt}= \left[-3g_2^2+2\lambda+6Y_t^2+2 {\rm
Tr}{\bf Y_e^\dagger Y_e} \right]\kappa -\frac{1}{2}\left[\kappa{\bf
Y_e^\dagger Y_e} + ({\bf Y_e^\dagger Y_e})^T\kappa\right]\ ,
\label{rg1}
\eea where $t=\log \mu$, and $g_2,\lambda, Y_t, {\bf Y_e}$ are the
$SU(2)$ gauge coupling, the quartic Higgs coupling, the top Yukawa
coupling and the matrix of Yukawa couplings for the charged leptons,
respectively. The last term of eq.(\ref{rg1}) is the most important
one for our purposes, since it modifies the texture of ${\cal
M}_{\nu}$.  It is worth noticing that  the
modification of a mass eigenvalue is always proportional to the mass
eigenvalue itself. 
In the MSSM case, things are very similar but with an important difference.
Namely, the term that modifies the texture in the RGEs has the 
same form as in eq.(\ref{rg1}) but with  coefficient $+1$ instead
of $-\frac{1}{2}$. Moreover, in the MSSM the ${\bf Y_e}$ couplings 
are $1/\cos \beta$ larger than the SM ones. All this implies that
the effect of the RGEs in the supersymmetric case is 
$2/\cos^2 \beta=2(1+\tan^2\beta)$ times larger (for $\tan\beta=2$ this 
already represents one order of magnitude).
It should be mentioned that in the supersymmetric case there are two stages
of running: from $\Lambda$ to $M_{SUSY}$ with the MSSM  RGEs, and
from $M_{SUSY}$ to $M_Z$ with the  SM ones (the latter is normally 
much less important than the former).

\vspace{0.2cm}
In order to study the quantitative effect of the RGEs on the mass splittings
and mixing angles, it is convenient to consider separately the following
three possible scenarios \cite{altfer}
\bea
{\rm A} & : & |m_3| \gg |m_{1,2}|, \nonumber\\
{\rm B} & : & |m_1|\sim |m_2| \sim |m_3|,\\
{\rm C} & : & |m_1|\sim |m_2| \gg |m_3|\nonumber .
\label{abc}
\eea
In case A, radiative corrections are generically not dangerous.
The reason is that, as stated before, the mass eigenvalues renormalize
proportionally to themselves, i.e. $\Delta_{RGE}\ m_i = (K_0 +K_i)m_i$, 
where $K_0$ is the universal contribution for all the eigenvalues and
$|K_i|\ll 1$. Thus, unless $m_{1,2}^2\gg\Delta m^2_{12}$, the running
cannot spoil the initial smallness of the solar 
mass-splitting (this is in particular the case of a hierarchical spectrum
$m_{1}^2\ll m_{2}^2\ll m_{3}^2$). 
Roughly speaking, the mass splittings generated
radiatively get larger than the allowed range of eq.~(\ref{VO}) for
$m_{1,2}^2\sim 10^{-4}$ eV$^2$, although the precise value depends
on several details, in particular on the values of the mixing angles.

On the other hand, case B [cosmologically relevant for
$m_i=\cal{O}$(eV)], has been shown to be inconsistent with the VO solution 
in refs.~\cite{EL,cein1,cein2}. 
Namely the mass splittings $\Delta m^2_{ij}$ generated 
through the running are several orders of magnitude larger than 
the required VO splitting, even for $\Lambda$ very close to $M_Z$.
According to the previous discussion, the supersymmetric case
works even worse.
The only way-out would be an extremely artificial
fine-tuning between  the initial values of the mass splittings 
(and mixing angles) and the effect of the RG running, something
clearly unacceptable.

Finally, the impact of the radiative corrections on a spectrum
of the type C has not been considered yet in the literature.
Since in this case the large $\Delta m^2_{sol}\ll\Delta m^2_{atm}$
hierarchy forces $m_{1,2}^2\gg\Delta m^2_{12}$, one can expect
important radiative effects.
The analysis of this scenario is performed in section 2, where
we study in two separate subsections the possibilities that 
$m_1$ and $m_2$ have equal or opposite signs. The conclusions
are presented in section 3.

\section{The case $m_1^2\sim m_2^2\gg m_3^2$}

As explained in the Introduction, radiative corrections
play an important r\^ole when $m_1^2\sim m_2^2\gg m_3^2$, case in which
the mass splitting relevant for solar oscillations is the one between the
heavier neutrinos. In this framework, radiative corrections 
can actually make $\Delta m_{12}^2 \gg \Delta m_{sol}^2$ in contradiction with
observations. This effect will be stronger the heavier is the overall
neutrino mass scale: the most conservative case thus corresponds to $m_1^2\sim
m_2^2\sim \Delta m_{atm}^2$ and $m_3^2\sim 0$ (masses smaller than this
cannot accommodate atmospheric oscillations of the required frequency).

The rationale is then the following: at some high-energy scale $\Lambda$
one assumes that new physics generates a dimension-5 operator leading 
to non-zero neutrino masses and fixes ${\cal M}_\nu(\Lambda)$ such that,
with good approximation $m_1^2=m_2^2$ and $m_3^2=0$. The most important
radiative corrections to this tree-level masses are proportional to
$\ln(\Lambda/M_Z)$ and can be included using standard RG techniques, that
is, running ${\cal M}_\nu$ down from $\Lambda$ to $M_Z$ using the relevant
RGEs. The latter depends on what is the effective theory below $\Lambda$.
As we said, we consider two cases: SM and MSSM, and
the RGEs relevant for these two effective theories can be found e.g. in
ref.~\cite{Babu}.

The analytical integration of the RGEs is straightforward in the
leading-log approximation and the additional simplification that 
$m_3\sim 0$ at all scales permits us to concentrate on the two other masses
alone. The results are qualitatively different
depending on the relative sign between $m_1$ and $m_2$ and we consider the
two cases separately in the next subsections.

\subsection{$m_1\simeq m_2$}

After integration from $\Lambda$ to $M_Z$, the radiatively corrected ${\cal
M}_\nu(M_Z)$ has eigenvalues which in first approximation are given by
\bear{cl}
\label{eigensame}
m_1 = & m_\nu,\vspace{0.2cm}\\
m_2 = & m_\nu \left[ 1 + 2 \epsilon_\tau (1-c_1^2c_2^2)
\right],\vspace{0.2cm}\\
m_3 = & 0.
\eear
These expressions include the leading-log radiative corrections to the mass
differences
and are obtained under the approximation that the initial 1-2 mass
splitting
is zero. In eq.(\ref{eigensame}), the family-universal renormalization
effect (not important for our discussion) has been absorbed in $m_\nu$,
which is fixed to give the proper value for
$\Delta m_{31}^2\sim \Delta m_{atm}^2$. The $\theta_1$, $\theta_2$, 
$\theta_3$ angles have been kept as free parameters.
Our numerical results are 
always obtained integrating numerically the RGEs and confirm that the
analytical expressions we write represent an excellent approximation. 
For our numerics we choose both the lower and upper limits of the allowed
range for $\Delta m_{atm}^2$, thus fixing 
$m_\nu=2.2\times 10^{-2}\, {\rm eV}$ or $0.1\, {\rm eV}$ (corresponding to
$\Delta m_{31}^2=
5\times 10^{-4}\, {\rm eV}^2$ and $10^{-2}\, {\rm eV}^2$, respectively).

The parameter $\epsilon_\tau$ depends on what is the effective low-energy
theory below $\Lambda$ \cite{cein1,cein2}:
\bea
\label{eps1}
\epsilon_\tau=\frac{h_\tau^2}{32 \pi^2}\log\frac{\Lambda}{M_Z}
\;\;\;\;({\rm SM}),
\eea
\bea
\label{eps2}
\epsilon_\tau=\frac{h_\tau^2}{32
\pi^2}\left[-\frac{2}{\cos^2\beta}\log\frac{\Lambda}{M_{SUSY}}+
\log\frac{M_{SUSY}}{M_Z}\right]
\;\;\;\;({\rm MSSM}),
\eea
where $M_{SUSY}$ sets the mass scale for the supersymmetric spectrum (we
take $M_{SUSY}\sim 1\, {\rm TeV}$). As usual, the size of $\epsilon_\tau$
grows logarithmically with the scale of new physics $\Lambda$ (a conservative
estimate we often make is to choose a low value $\Lambda=1\, {\rm TeV}$). 
Also, for
sufficiently large $\Lambda/M_Z$, the size of $\epsilon_\tau$ is enhanced
by a factor $2/\cos^2\beta=2(1+\tan^2\beta)$ in the MSSM with respect to
the SM (already a factor 10 for $\tan\beta=2$) so that radiative
corrections are more important in this case.

The typical size of $\epsilon_\tau$ is $\sim 8\times 10^{-7}$  ($\sim
8\times 10^{-6}$) for $\Lambda=10^3\, {\rm
GeV}$ ($\Lambda=10^{12}\, {\rm GeV}$) in the SM and $\sim -8\times 10^{-5}$
for $\Lambda=10^{12} \, {\rm GeV}$ in the MSSM with $\tan\beta=2$. According
to
Eq.~(\ref{eigensame}) this would lead to 
\bea
\frac{\Delta m_{sol}^2}{m_\nu^2}=\frac{\Delta m_{sol}^2}{\Delta m_{atm}^2}
=4\epsilon_\tau (1-c_1^2c_2^2),
\label{splitting}
\eea
too large compared with the observed value unless there is a cancellation
in $(1-c_1^2c_2^2)$, which requires $\sin^2 2\theta_{1,2}\sim 0$. This is
far from the best-fit values mentioned in the Introduction. Choosing
$\sin^2 2\theta_1\simeq 1$ and $\sin^2 2\theta_2\simeq 0$ we must conclude
that $\Delta m_{12}^2$ turns out to be too large for 
vacuum oscillations of solar neutrinos.
\begin{figure}[t]  
\centerline{
\psfig{figure=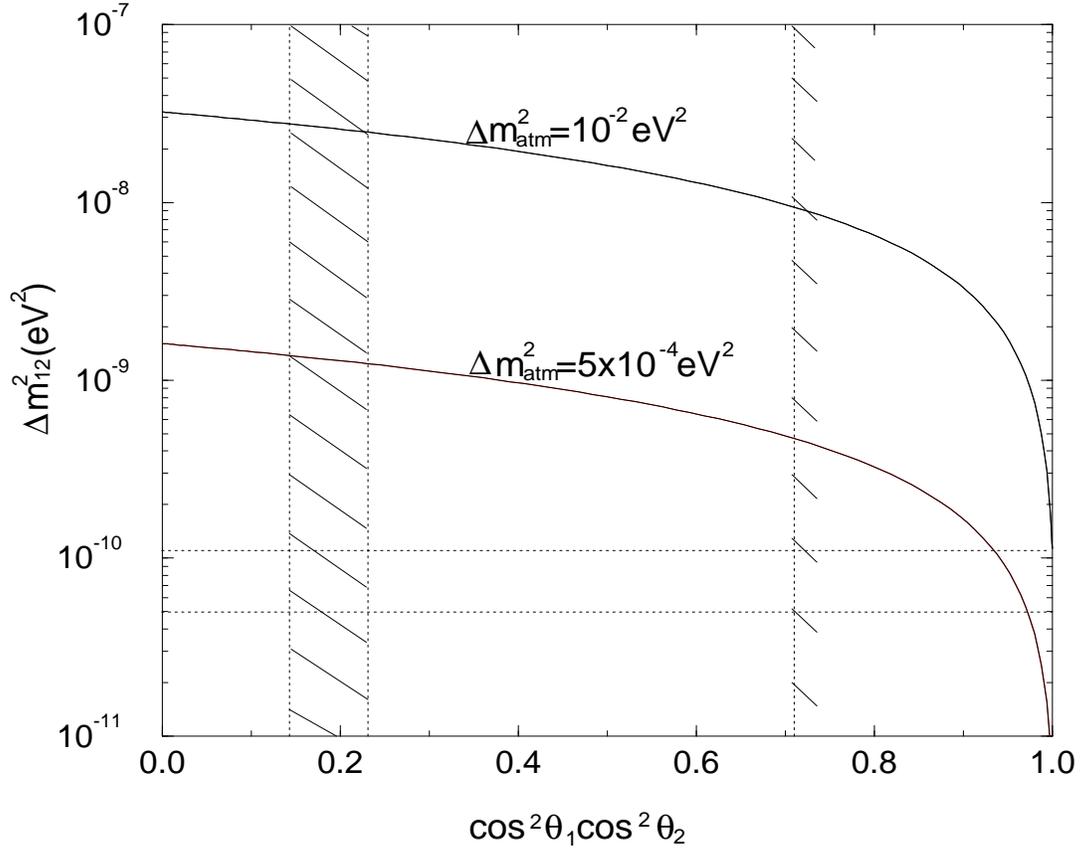,height=12cm,width=10cm,bbllx=4.cm,%
bblly=1.cm,bburx=16.cm,bbury=16.cm}}
\caption
{\footnotesize 
$\Delta m_{12}^2/{\rm eV}^2$ (solid lines) as a function of $c_1^2c_2^2$
for $\Delta m_{atm}^2=10^{-2}\, {\rm eV}^2$ (upper curve) and $5\times
10^{-4}\,
{\rm eV}^2$ (lower). The experimentally allowed region for $c_1^2c_2^2$ is
delimited by the dashed lines (the dashed regions are forbidden) and the
$\Delta m_{sol}^2$ needed for the VO solution is given by the range between
the dotted lines. This plot corresponds to the SM with $\Lambda=10^3\, {\rm
GeV}$.} 
\end{figure}

The precise results are given in Figure 1, which shows the predicted 
$\Delta m_{12}^2$ in ${\rm eV}^2$ (solid
lines) as a function of $c_1^2c_2^2$ for the SM case with $\Lambda=10^3\, 
{\rm GeV}$ and $\Delta m_{atm}^2=5\times 10^{-4}\, {\rm eV}^2$ (lower
curve) and $10^{-2}\, {\rm eV}^2$ (upper curve). The experimental
constraints on $\theta_{1,2}$ leave open the windows $0\leq c_1^2c_2^2
\leq 0.142$ and $0.232\leq c_1^2c_2^2 \leq 0.71$, as indicated.
The neutrino mass splitting required by VO
solar oscillations is marked by the horizontal dotted lines. As was
clear from the previous discussion, there is no overlapping between the
$\Delta m_{12}^2$ predicted and the $\Delta m_{sol}^2$ required. 
Indeed, $\Delta m_{12}^2$ is always much larger than the allowed range.
In the MSSM (or for larger $\Lambda$) the situation is even worse because 
in both cases $\Delta m_{12}^2$ increases significantly in the way
discussed above.

Let us turn in more detail to the mixing angles in this scenario. At the
scale $\Lambda$ one has some mixing angles $\theta_i$ which will be
different in general at the scale $M_Z$ after
radiative corrections to ${\cal M}_\nu$ have been included. At the same
level of approximation as in Eqs.~(\ref{eigensame}), the
eigenvectors of the perturbed neutrino mass matrix are of the form
\bea
V_1' = \frac{1}{\sqrt{\alpha^2+\beta^2}}
(\alpha V_1+\beta V_2),\;\;\;
V_2' = \frac{1}{\sqrt{\alpha^2+\beta^2}}(-\beta V_1+\alpha V_2),\;\;\;
V_3' = V_3,
\eea
where $V_i$ are the eigenvectors corresponding to ${\cal M}_\nu(\Lambda)$
\bea
V_1 = \left (
\begin{array}{c}
c_2c_3 \vspace{.1cm}\\
-c_1s_3-s_1s_2c_3  \vspace{.1cm}\\
s_1s_3-c_1s_2c_3
\end{array}
\right ), ~~
V_2 = \left (   
\begin{array}{c}
c_2s_3 \vspace{.1cm}\\ 
c_1c_3-s_1s_2s_3 \vspace{.1cm}\\
-s_1c_3-c_1s_2s_3 \end{array}
\right ), ~~
V_3 = \left (
\begin{array}{c}
s_2  \vspace{.1cm}\\ 
s_1c_2 \vspace{.1cm}\\
c_1c_2 \end{array}  
\right ).
\label{vecbim}
\eea
From this, we deduce that the relationships between $\theta_i(M_Z)$ and
$\theta_i(\Lambda)$ are 
\bear{cl}
\label{anglesame}
\sin^22\theta_1(M_Z)=&\sin^22\theta_1(\Lambda),\vspace{0.2cm}\\
\sin^22\theta_2(M_Z)=&\sin^22\theta_2(\Lambda),\vspace{0.2cm}\\
\sin^22\theta_3(M_Z)=&\sin^22\theta_3(\Lambda)+{\displaystyle
\frac{2}{(1+r)^2}}\left[
\sqrt{r}(1-r)\sin 4\theta_3(\Lambda)+2r\cos 4\theta_3(\Lambda)
\right],
\eear
where $r\equiv \alpha^2/\beta^2$. In leading-log approximation we have
\bea
\frac{\alpha}{\beta}\simeq \frac{s_1c_3+c_1s_2s_3}{s_1s_3-c_1s_2c_3},
\eea
with all angles evaluated at the scale $\Lambda$.

If we substitute this in (\ref{anglesame}) we find the simpler expression
\bea
\sin^22\theta_3(M_Z)=\frac{\sin^2\theta_2\sin^22\theta_1}{
(1-\cos^2\theta_1\cos^2\theta_2)^2}.
\eea
For the bimaximal mixing case 
($s_2\sim 0$, $c_1\sim s_1\sim 1/\sqrt{2}$) we end up with
$\sin^22\theta_3(M_Z)\sim 0$, which is not acceptable (observations require
$\sin^22\theta_3 \geq 0.67$).

In conclusion, the scenario $m_1\sim m_2\gg m_3$ is very contrived 
from the theoretical point of view. It is not natural to expect
in this framework the values of mass splittings and mixing angles which
are suggested by experiment. As mentioned in the Introduction, 
the only way-out would be an extremely artificial
fine-tuning between  the initial values of the mass splittings 
(and mixing angles) and the effect of the RG running. If one
insists on this possibility, starting for example with 
${\Delta m_{atm}^2=5\times 10^{-4}\ {\rm eV}^2}$,
$s_2\sim 0$, $c_1\sim s_1\sim 1/\sqrt{2}$, $\Lambda=10^3$ GeV 
(a conservative choice for the fine-tuning problem), one is forced to 
take the initial mass splitting and mixing angle within
the narrow ranges $|m_1^2-m_2^2|\sim (1.82\pm
0.02)\times 10^{-7}\, {\rm eV}^2$ and $\theta_3\sim\pi/2\pm 5.5\times
10^{-3}$ in order to compensate the effect of the RGEs and reproduce the
required 
pattern of masses and mixings at $M_Z$ [these numbers cannot
be extracted from the previous eq.(\ref{splitting}), as in this case the
approximation of initial degenerate eigenvalues does not hold].
One cannot certainly expect such a conspiracy 
between totally unrelated effects. 
If one slightly separates from these narrow ranges
the low-energy mass splitting would be much larger than the 
required one. Of course,  as $\Delta m_{atm}^2$ or
$\Lambda$ are raised, or one goes to the supersymmetric case, 
the fine-tuning becomes much stronger.

Finally, it is interesting to note that for sizeable values of the 
cut-off ($\Lambda\simgt 10^{12}\ {\rm GeV}$) and/or a supersymmetric
scenario, the values of $\Delta m_{12}^2$ are naturally 1-3 orders
of magnitude larger than those represented in Fig.1, falling in the
small-angle MSW range ($3\times10^{-6}\ {\rm eV}^2 < 
\Delta m^2_{sol} < 10^{-5}\ {\rm eV}^2$). This is appealing since, 
as has been discussed
in this section, starting with $\theta_{1,2}$ mixing angles in agreement
with experiment ($s_2\sim 0$, $c_1\sim s_1\sim 1/\sqrt{2}$) the
RGEs drive $\sin^22\theta_3(M_Z)\sim 0$, independently of its initial
high-energy value, see eq.(\ref{anglesame}). 
This is exactly what is needed for a successful small-angle MSW solution 
to the solar neutrino problem.

\subsection{$m_1\simeq -m_2$}

In this case, the neutrino mass eigenvalues at $M_Z$ are, in leading-log
approximation
\bear{cl}
\label{eigendif}
m_1 = &
m_\nu\left[1+2\epsilon_\tau(s_1s_3-c_1s_2c_3)^2\right],\vspace{0.2cm}\\
m_2 = & -m_\nu \left[ 1 + 2 \epsilon_\tau (s_1c_3+c_1s_2s_3)^2\right],
\vspace{0.2cm}\\
m_3 = & 0,
\eear
with $\epsilon_\tau$ as given by Eqs.~(\ref{eps1}) and (\ref{eps2}). The
mixing angles are, in first approximation, equal at $M_Z$ and $\Lambda$.
We fix again $m_\nu\sim\sqrt{\Delta m_{atm}^2}$. In order not to 
spoil the size of the required solar mass splitting, the radiative
corrections should generate $\Delta
m_{12}^2\leq\Delta m_{sol}^2$.
The prediction from (\ref{eigendif}) is
\bea
\frac{\Delta m_{12}^2}{m_\nu^2}=\frac{\Delta m_{12}^2}{\Delta m_{atm}^2}
=4\epsilon_\tau\left[(\cos^2\theta_1 \sin^2\theta_2
-\sin^2\theta_1)\cos 2\theta_3-\sin 2\theta_1\sin\theta_2\sin 2\theta_3
\right].
\eea
Getting a sufficiently small number for this quantity requires some 
(in general delicate) correlation
between the mixing angles, in such a way that
\bea
\label{cancel}
\tan 2\theta_3\simeq \frac{\cos^2\theta_1 \sin^2\theta_2
-\sin^2\theta_1}{\sin\theta_2\sin 2\theta_1}.
\eea
It is remarkable that the bimaximal values of the mixing angles
($\sin^22\theta_1\sim\sin^22\theta_3\sim 1$ and $\sin^22\theta_2\sim 0$)
do satisfy (\ref{cancel}).

\begin{figure}
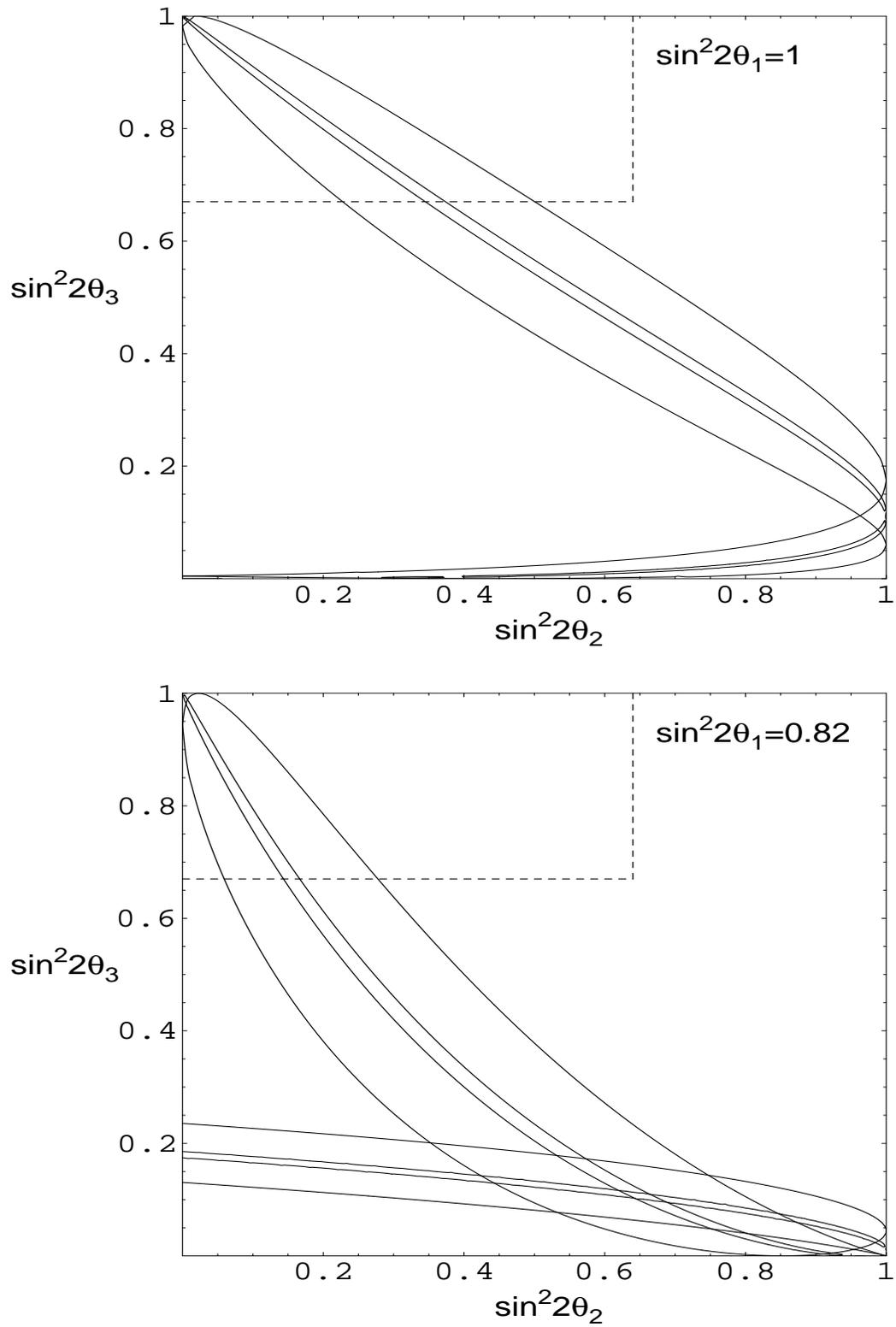

\centerline{\vbox{
\psfig{figure=vocein2a.ps,height=10cm,width=10cm,bbllx=7.cm,%
bblly=5.cm,bburx=15.cm,bbury=15.cm}
\psfig{figure=vocein2b.ps,height=10cm,width=10cm,bbllx=7.cm,%
bblly=5.5cm,bburx=15.cm,bbury=15.5cm}}}
\caption
{\footnotesize Allowed regions (in which $\Delta m_{12}^2\leq 1.1\times
10^{-10}\, {\rm eV}^2$) in the plane ($\sin^22\theta_2,\sin^22\theta_3$)
for the SM case for $\sin^22\theta_1=1$ (upper plot) and 0.82 (lower). 
The area between the outermost (innermost) lines corresponds to
a cut-off scale $\Lambda=10^3\, {\rm GeV}$ ($10^{12}\, {\rm GeV}$).
The dashed lines 
delimit the experimentally allowed values for
$\sin^22\theta_2$ and $\sin^22\theta_3$. } 
\end{figure}

Figures 2a,b show, for the SM case, the regions in the plane
$(\sin^22\theta_2,\sin^22\theta_3)$
where the correlation (\ref{cancel}) takes place, giving $\Delta
m_{12}^2\leq 1.1\times 10^{-10}\, {\rm eV}^2$ (the upper limit on $\Delta
m^2_{sol}$).
The width of these allowed regions is controlled by $\Delta
m_{sol}^2/(\epsilon_\tau\Delta
m_{atm}^2)$. The larger $\epsilon_\tau\Delta
m_{atm}^2$ (or the smaller $\Delta m^2_{sol}$), the thinner these
regions get
[because a more delicate
cancellation must take place in (\ref{cancel})].
In figure~2a we have fixed $\sin^22\theta_1=1$ and $\Delta m_{atm}^2=5\times
10^{-4}\, {\rm eV}^2$, and we give the allowed areas for the two choices 
$\Lambda=10^3\, {\rm GeV}$ (thick region, delimited by the outermost lines)
and $10^{12}\, {\rm GeV}$ (thin
region). If we choose $\Delta m_{atm}^2=10^{-2}\, {\rm eV}^2$ instead, the
two regions would shrink significantly and will be somewhere inside the thin
 region shown for $\Lambda=10^{12}\, {\rm GeV}$.
The dashed lines delimit the allowed region for the two mixing angles
$\theta_2$ and $\theta_3$ ($0\leq \sin^22\theta_2\leq 0.64$ and
$0.67\leq\sin^22\theta_3\leq 1$). 
Figure~2b corresponds to the case $\sin^22\theta_1=0.82$ (the
lower experimental limit) and the same values of other parameters as in
figure~2a.  
The results are similar except for a shift towards smaller $\sin^22\theta_2$
values in the region of interest. Note in particular that the upper limit
$\sin^22\theta_2\sim 0.64$ is
never reached in this scenario.
We see that in the most conservative case,
with $\Delta m_{atm}^2=5\times 10^{-4}\, {\rm eV}^2$ and $\Lambda =10^3\, {\rm 
GeV}$, a significant portion of parameter space could accommodate a
$\Delta
m_{12}^2$ of the right order of magnitude (including the bimaximal mixing
solution). It is interesting to note that inside this region, starting with 
degenerate $m_1,\ m_2$ would lead to a correct $\Delta m_{sol}^2$ at low
energy, thus providing a dynamical origin for this small number.
Notice however that as soon as $\Delta m_{atm}^2$ or
$\Lambda$ are raised the required fine-tuning becomes much stronger.
This occurs in particular if the lower bound 
$\Delta m_{atm}^2=5\times 10^{-4}$ that we have used is
increased according to the analyses of the most recent data 
\cite{SK99}.

The situation is worse in the MSSM case. Roughly speaking, for $\tan\beta=3$
radiative corrections are 20 times larger than in the SM (with the same
$\Lambda$). The cancellation between mixing
angles in (\ref{cancel}) is thus much more delicate in the supersymmetric
case, as expected.

\section{Conclusions}

The vacuum oscillation (VO) solution to the solar neutrino problem
requires an extremely small mass splitting, $\Delta m^2_{sol}\simlt
10^{-10}\ {\rm eV}^2$. We have studied in this paper under which
circumstances this smallness (whatever its origin) is or is not
spoiled by radiative  corrections, in particular by the running  of
the renormalization group equations (RGEs) between the scale at which
the effective neutrino mass matrix is generated  ($\Lambda$) and low
energy. We consider the cases where the effective
theory below $\Lambda$  is the Standard Model (SM) or the Minimal
Supersymmetric Standard
Model (MSSM).  The results depend dramatically on the type of neutrino
spectrum.  In particular, if $m_1^2\ll m_2^2\ll m_3^2$,  radiative
corrections are always relatively small and do not cause any
significant change in the splittings. On the other hand, if $m_1^2\sim
m_2^2\sim m_3^2$, radiative corrections always induce mass
splittings that are several orders of magnitude larger than  the
required $\Delta m^2_{sol}$. Hence, this type of spectrum is
not plausible for the VO solution.  The only way-out would be an
extremely artificial fine-tuning between  the initial values of the
mass splittings  (and mixing angles) and the effect of the RG running,
something clearly unacceptable. 

Most of the paper is devoted to the third possible type of spectrum, 
$m_1^2\sim m_2^2\gg m_3^2$, which requires $m_{1,2}^2\sim
\Delta m^2_{atm}$ (or larger). Here again, the radiatively generated
splittings are in general too large, making the scenario unnatural.
As a general rule, this gets worse as $\Delta m_{atm}^2$ or
$\Lambda$ grow. Also, the supersymmetric scenario works worse
than the SM one, especially as $\tan \beta$ incresases.
More precisely, if $m_1$ and
$m_2$ have equal signs, the RGE-induced splittings are always
too large, even for the most favorable case. In addition, 
the solar mixing
angle is driven by the RGEs to very small values,
$\sin^22\theta_3(M_Z)\sim 0$, which is incompatible with the VO
solution. It is however worth noticing that such a small
angle is what is needed for a successful small-angle MSW solution 
to the solar neutrino problem. Moreover, for 
$\Lambda\simgt 10^{12}\ {\rm GeV}$ and/or for the MSSM scenario
the values of $\Delta m_{12}^2$ may fall naturally in the
small-angle MSW range,  $\sim 10^{-5}\ {\rm eV}^2$.

If $m_1$ and $m_2$ have opposite signs, the results are
analogous, but now the splitting generated by the RGEs can vanish if
the mixing angles are correlated in a particular way (which remarkably
is always satisfied by the exact bimaximal case). 
This correlation or
tuning of parameters is acceptable in the SM scenario, 
provided the cut-off scale
$\Lambda$ is not much larger than $\sim 1\, {\rm TeV}$ and if $\Delta
m_{atm}^2$ is in the low side of its experimentally preferred range ($\sim
5\times 10^{-4}\, {\rm eV}^2$). Interestingly, this could provide a 
dynamical origin for the smallness of $\Delta m^2_{sol}$.
For larger $\Lambda$ and/or $\Delta
m_{atm}^2$ (or equivalently, for the MSSM scenario) radiative 
corrections grow in size
and the required tuning of mixing angles becomes quickly unacceptable.
This occurs in particular if the lower bound 
$\Delta m_{atm}^2=5\times 10^{-4}$ that we have used is
increased according to the most recent data analyses \cite{SK99}.

In conclusion, apart from the mentioned small windows, a
completely hierarchical spectrum of neutrinos (i.e. as the
spectrum of quarks and charged leptons), $m_1^2\ll m_2^2\ll m_3^2$,
seems to be the only plausible one for the VO solution to the
solar neutrino problem.

\section*{Acknowledgements}

This research was supported in part by the CICYT
(contract AEN98-0816). 
A.I. and  I.N. thank the CERN Theory Division for hospitality. 



\begin{thebibliography}{99}
%
\bibitem{range} R.~Barbieri et al., \JHEP 9812 98 017; G.L.~Fogli, E.~Lisi,
A.~Marrone and G.~Scioscia,
{\sl Phys.~Rev.}~{\bf D59}:033001 (1999), [hep-ph/9904465].
%
\bibitem{SK} Y.~Fukuda et al.,
Super-Kamiokande Collaboration, \PLB 433 98 9; \PRL 81 98 1562; 
S.~Hatakeyama et al., Kamiokande Collaboration,
\PRL 81 98 2016.
%
\bibitem{CL} First ref. in \cite{range}; A.~Strumia, \JHEP 9904 99 026.
%
\bibitem{sym}
Y. L. Wu,
[hep-ph/9810491], [hep-ph/9901245] and [hep-ph/9901320]; C. Wetterich,
\PLB 451 99 397; R.~Barbieri, L.J.~Hall, G.L.~Kane, and G.G.~Ross,   
[hep-ph/9901228]; M.~Tanimoto, T.~Watari, T.~Yanagida, [hep-ph/9904338].
%
\bibitem{eff} S. Weinberg, \PRL 43 79 1566.
%
\bibitem{Babu} K. Babu, C. N. Leung and J. Pantaleone, \PLB 319 93 191.
%
%
\bibitem{altfer} G. Altarelli and F. Feruglio,
\PLB 439 98 112, \JHEP 9811 98 021 , and
\PLB 451 99 388.
%
\bibitem{EL} J.
Ellis and S.
Lola, [hep-ph/9904279].
%
\bibitem{cein1} J.A. Casas, J.R. Espinosa, A. Ibarra and I. Navarro,
[hep-ph/9904395].
%
\bibitem{cein2} J.A. Casas, J.R. Espinosa, A. Ibarra and I. Navarro,
[hep-ph/9905381].
%
\bibitem{SK99} K. Scholberg, [hep-ex/9905016].
\end{thebibliography}
\end{document}